\documentclass{aa}
\usepackage{booktabs}
\usepackage{array}
\usepackage{supertabular}
\usepackage{siunitx}
\usepackage{subfig}
\usepackage{txfonts}
\usepackage{hyperref}
\usepackage{amstext}

\begin{document}

    \title{Possible hadronic origin of TeV photon emission from SNR G106.3+2.7}

    \author{Chuyuan, Yang\inst{1}
    \and
    Houdun, Zeng \inst{2,3}
    \and
    Biwen, Bao \inst{3,4}
    \and
    Li, Zhang \inst{3}
        }
    \institute{Yunnan Observatories, Chinese Academy of Sciences, Kunming 650011, China % 1
  %%  \newline e-mail: chyy@ynao.ac.cn
    \and
    Key Laboratory of Dark Matter and Space Astronomy, Purple Mountain Observatory, Chinese Academy of Sciences, Nanjing 210034, China % 2
    \and
    Department of Astronomy, Key Laboratory of Astroparticle Physics of Yunnan Province, Yunnan University, Kunming 650091, China % 3
    \and
    Key Laboratory of Statistical Modeling and Data Analysis of Yunnan Province, Yunnan University, Kunming 650091, China % 4
   \newline e-mail: lizhang@ynu.edu.cn
    }

    \date{Received ; accepted }
    \titlerunning{Possible hadronic origin of emission from SNR G106.3+2.7}
    % \abstract{}{}{}{}{}
    % 5 {} token are mandatory

    \abstract
     {
     \textbf{Context.} Recently, HAWC, AS$\gamma$, and LHAASO experiments have reported the gamma-ray spectrum of supernova remnant (SNR) G106.3+2.7 above 40 TeV, indicating that SNR G106.3+2.7 is a promising PeVatron candidate. However, the origin of the gamma-ray spectrum is still debated. Thus, a dedicated theoretical model with self-consistent descriptions is required to decipher the properties of the gamma-ray spectrum for this specific source.

     \textbf{Aims.} We construct a theoretical model to explain the multiband photon emission from the PeVatron SNR G106.3+2.7.

     \textbf{Methods.} In our model, the acceleration and propagation of particles from the Bohm-like diffusion region inside the SNR to the Galactic diffusion region outside the SNR are described through nonlinear diffusive shock acceleration (NLDSA). The main content of our NLDSA model is solving the hydrodynamic equations numerically for gas density, gas velocity, and gas pressure and the equation for the quasi-isotropic particle momentum distribution. The consequent multiband nonthermal emission stems from two different regions, namely the acceleration region and the escaping region.

     \textbf{Results.} Our model is capable of explaining the multiband photon emission via the dominant synchrotron radiation of the electrons accelerated inside the SNR. The photons with energy of $\gtrsim$ GeV are naturally produced by the protons inside and outside the SNR. Moreover, photons in the energy range of $\sim 1 - \sim 100$ TeV are due to the interaction of escaped protons with dense molecular clouds.

     \textbf{Conclusions.} For photons with energy $E_\gamma \gtrsim $ 1 GeV from SNR G106.3+2.7, our results here favor a hadronic origin, where the photons in the energy range of  $\sim 1$ GeV to $\sim 1$ TeV are produced inside the SNR through proton-proton interaction, while photons with $E_\gamma \gtrsim 1$ TeV originate from the interaction of escaped protons with a dense molecular cloud.
    }

    \maketitle
    %
    %________________________________________________________________

\section{Introduction}\label{sect:intro}
Although particles in supernova remnants (SNRs) can be efficiently accelerated up to $\text{about }$PeV energy through the diffusive shock acceleration (DSA) mechanism \citep{Ptuskin2010}, no direct observational evidence for the acceleration of PeV Galactic cosmic rays in SNRs has been reported so far. As cosmic rays (CR) are deflected by Galactic magnetic fields, it is difficult for us to trace their sources. However, GeV-TeV gamma-ray emission can be produced through relativistic protons interacting with gas and/or a molecular cloud (MC) near the SNRs \citep[e.g.,][]{Aharon2001, Gabici2007, Ohira2011}. At least two SNRs (IC 443 and W 44) appear to have the characteristic pion bump feature in the gamma-ray spectra \citep{Ackermann2013}, helping us to understand the particle acceleration mechanism.

A large number of SNRs have been detected in the gamma-ray band, but none of these SNRs has been shown to emit gamma-rays to hundreds of TeV, that is, a PeV particle accelerator (or PeVatron) has not yet been confirmed experimentally.  Recently, HAWC, AS$\gamma$, and LHAASO experiments \footnote{HAWC: The High-Altitude Water Cherenkov Gamma-Ray Observatory. AS$\gamma$: Tibet air shower experiment. LHAASO: Large High Altitude Air Shower Observatory.} have reported a gamma-ray spectrum of SNR G106.3+2.7 above 40 TeV, indicating that SNR G106.3+2.7 is a promising PeVatron candidate \citep{Albert2020,Cao2021}. Observation of gamma-ray emission from SNR G106.3+2.7 above 10 TeV, even up to a few 100 TeV, has also been reported \citep{Amene2021}, indicating that the very high-energy (VHE) gamma-ray emission above 10 TeV is well correlated with an MC rather than with the pulsar PSR J2229+6114. This morphological feature favors a hadronic origin of the VHE emission.

Recently,  \citet{Xin2019} argued that the pure leptonic model for the $\gamma$-ray emission can be ruled out based on the X-ray and TeV data from SNR G106.3+2.7. According to the current understanding of acceleration and escaping processes, \citet{Bao2021} explored a new scenario to explain the $\gamma$-ray spectrum for this source. They considered the hadronic $\gamma$-ray spectrum dominating the emission above 10 TeV to be due to the accelerated protons escaping and interacting with MCs.

 The CR particles in SNRs can been accelerated through the DSA mechanism \citep[e.g.,][]{Kang2006, Caprioli2008,Zirakashvili2012,Telezhinsky2012a, Ferrand2014}. They escape from SNRs at the so-called free escape boundary \citep[e.g.,][]{Ellison2011,Telezhinsky2012b, Kang2013}.

In this paper, a theoretical model is developed to explain the multiband emission from SNR G106.3+2.7, in particular, GeV to $\sim100$ TeV  emission. In our model, the particles are accelerated inside an SNR through DSA with Bohm-like type diffusion. Then the accelerated particles propagate outward into a different region with Galactic diffusion outside the free escape boundary. These escaped particles can produce $\gtrsim 100$ TeV photons through $\pi^0$ decay when they encounter dense MCs. Therefore, the multiband emission from the SNR G106.3+2.7 can be consistently reproduced.

\section{Modeling multiband emission from SNR G106.3+2.7}\label{mod}

\subsection{Basic properties of observed multiband emission}\label{per}

Supernova remnant G106.3+2.7 has been observed at radio, X-ray, and $\gamma$-ray bands. Its basic features are as follows:
(i) At the radio band, a brighter head and an extended tail region are evident. The overall spectral index is $\sim0.57$, and the tail region has a slightly steeper spectrum than that in the head region \citep{Pineault2000}.
(ii) At the X-ray band, the observational data of Suzaku targeting the SNR and adjacent pulsar PSR J2229+6114 indicate that the diffusive X-ray emission is nonthermal and originates from leptonic synchrotron emission with a photon index of $\sim 2.2$ \citep{Fujita2021}.
(iii) At the $\gamma$-ray band, a detailed analysis of GeV emission near VER J2227+608 with ten years of Fermi-LAT Pass 8 data was performed \citep[]{Xin2019}. Significant TeV $\gamma$-ray emission from the elongated radio extension of SNR G106.3+2.7 was detected by VERITAS \citep[]{Acciari2009}. In this region, emission with tens of TeV was also detected by Milagro \citep[]{Abdo2007,Abdo2009}. Multi-TeV emission coincident with G106.3+2.7 has also been detected by HAWC, and the joint VERITAS-HAWC spectrum is well fitted by a power law from 900 GeV to 180 TeV \citep[]{Albert2020}. Recently, $\gtrsim 100$ TeV photons from the direction of SNR G106.3+2.7 were detected by AS$_\gamma$ experiment and LHAASO \citep[]{Amene2021,Cao2021}. Observationally, the spectral energy distribution between $\sim 1$ GeV and $\sim 10$ TeV at first appears as an increasing trend, then decreases sharply, extending to $\gtrsim 100$ TeV. Therefore, SNR G106.3+2.7 is a new potential Galactic PeVatron.

The age of SNR G106.3+2.7 still remains uncertain, with several speculations: $\sim 1$ kyr \citep[][]{Albert2020,Bao2021}, or about $10^4$ yr in comparison with the characteristic age of PSR J2229+6114. Here we adopt an age of 2000 yr to match the radius ($\sim 5.7$ pc at a distance of 800 pc) of the SNR evolution with reasonable parameters (the explosion energy $E_{\rm sn}=10^{50}$ erg and ejecta mass $M_{\rm ej}=1.4 M_\odot$). With the adiabatic model of \citet{Kothes2001} ($n_0=1.9\times10^{24}R_f^{-5} t^2 E_{\rm sn}$, and $R_f$ is the shock radius), the ambient medium number density is estimated as $n_0\sim0.2$ cm$^{-3}$.

\subsection{Modeling and results}

As mentioned in \S \ref{sect:intro},  we assumed that the multiband nonthermal emission stems from two different regions, namely the acceleration region and the escaping region. These two regions, separated by the free escape boundary $R_{\rm esc}$, retain different diffusion coefficients. Inside the radius of $R_{\rm esc}$, the actual diffusion resembles a Bohm-like type, while outside the radius, Galactic diffusion dominates. For this purpose, the numerical nonlinear diffusive shock acceleration (NLDSA) model developed in \citet{Zirakashvili2012} (hereafter ZP2012 model) was adopted, as it could naturally extend from the Bohm-like diffusion region to the Galactic diffusion region \citep{Yang2015,Tang2015}.

The main content of our NLDSA model is solving the hydrodynamic equations numerically for gas density $\rho(r, t)$, gas velocity $u(r, t)$, and gas pressure $P_{\rm g} (r , t) $ and the equation for the quasi-isotropic particle momentum distribution $N(r, t, p) $ (see Equations (1)-(4) of ZP2012). We extended the ZP2012 model to the Galactic diffusion region and explained TeV emission from two TeV sources in our previous work \citep{Tang2015}.

In the acceleration region, the magnetic field above the radius $R_{ c}$ of the contact discontinuity (CD) is given by ZP2012, and the pressure and energy flux of the magnetic field are introduced in the dynamical equation \citep{Zirakashvili2014}. In the ZP2012 model, the coordinate dependences of the magnetic field are shown as follows:
\begin{equation}
 B(r,t)=B_0 \frac{\rho}{\rho_0}\sqrt{\frac{\dot{R_{ f}^2}}{M_{A}^2V_{\rm A0}^2}+1}, r > R_{ c}\;,
\end{equation}
where $\rho_0$ and $V_{A0}=B_0/\sqrt{4\pi\rho_0}$ are the gas density and the Alfv$\acute{e}$n velocity of the circumstellar medium; $M_{ A}$ determines the value of the magnetic field amplification,  and for young SNRs $M_{ A}$ , it is about $23$ (ZP2012); $R_{f}$ is the radius of the forward shock (FS).
However, in the region of $\le R_{\rm c}$, the magnetic field is
\begin{equation}
B(r,t)=\sqrt{4\pi\rho_m}\frac{|\dot{R_b}-u(r_m)|}{M_A}\left\{
\begin{array}{ll}
1, & r<r_m, \\
\frac{\rho}{\rho_{ m}}, & r_{ m} <r<R_{ b},\\
\frac{\rho(R_b+0)}{\rho_m},& R_b<r<R_c.
\end{array} \right.
\end{equation}
Here $r_m<R_b$ is the radius where the ejecta density has a minimum and equals $\rho_m$. The particle diffusion in the acceleration region is assumed to be a Bohm-like type, and the diffusion coefficient is given by
$D=\eta_B D_{ B}$, where $\eta_B =2$ represents the possible deviations of diffusion coefficient from the typical Bohm value in a high-velocity shock, $D_{B}=vpc/3qB$ is the Bohm diffusion coefficient (also see ZP2012 for details). %in a radius of $1.2R_f$ (inside SNRs) with

\begin{figure}
        \centering

                \includegraphics[width=8.7cm]{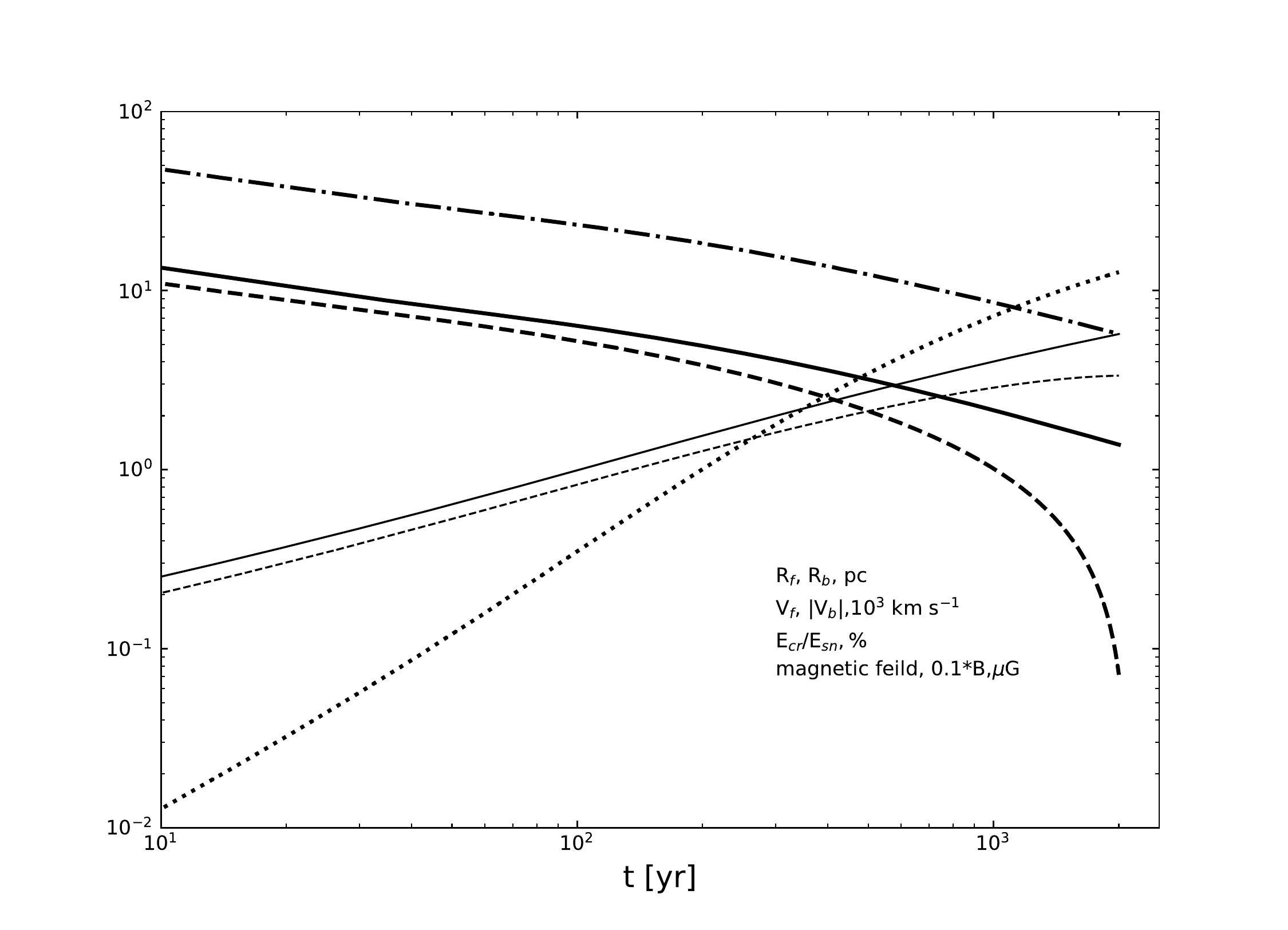}

        \caption{Dependences on time of the forward-shock radius $R_f$ (thin solid line), the reverse-shock radius $R_b$ (thin dashed line), the forward-shock velocity $V_f$ (thick solid line), the reverse-shock velocity $|V_b|$ (thick dashed line),  the ratio of CR energy and energy of supernova explosion $E_{cr}/E_{sn}$ (thick dotted line),  and the magnetic field strength (thick dash-dotted line) in the shock radius ($r=R_f$).}
        \label{fig:1}
\end{figure}

In the escaping region, the magnetic field is assumed to be the classical Galactic magnetic field, and the diffusion coefficient can be expressed as
\begin{equation}
D_g(E_p)=10^{28}\chi \left(\frac{E_p}{10 {\rm GeV}}\right)^{0.5} \rm cm^2s^{-1}\;,
\end{equation}
where $E_p$ is the accelerated proton energy, and the modification factor of $\chi=10^{-3}$ is for slow diffusion and is much lower than that of the averaged Galactic diffusion ($\chi \sim 1$), which indicates that the diffusion is suppressed by a factor of 100 or more near SNRs \citep{Fujita2009}. According to $l=2\sqrt{tD_g(E_p)}$ \citep{Atoyan1995}, we have $l\sim5$ pc for a 100 TeV in a diffusing particle of 2000 yr. The escaped particles with a few hundred TeV remain in the interacting region with the nearby MC.

In our calculations, the SNR shock propagates in the ambient medium with an average hydrogen number density of $ n_0=0.2 \rm cm^{-3}$, a magnetic field strength of $B_0=5 \mu\rm G$, a temperature of $T=10^4$ K, an index of the ejecta velocity distribution of $k=7$, and an initial forward-shock velocity of $2.9\times 10^4 \rm\  km\  s^{-1}$.  The injection efficiency of thermal protons and electrons is assumed to be $\eta_f=\eta_b=10^{-4}$ and $\eta_f=\eta_b=10^{-7}$ at FS ($f$) and RS ($b$), respectively.

\begin{figure}
        \centering
        \begin{minipage}{8cm}
                \includegraphics[width=8.7cm]{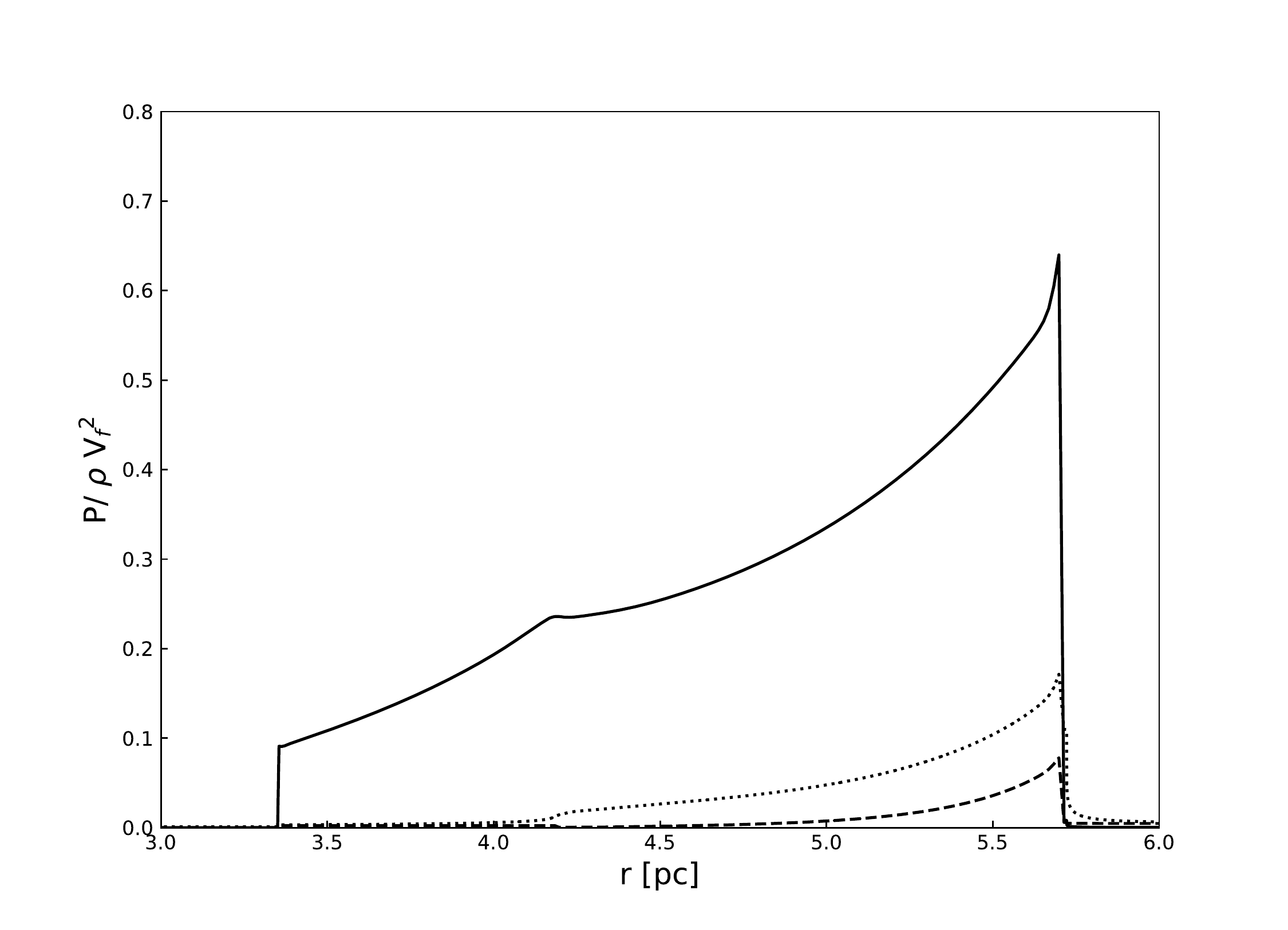}
        \end{minipage}
        \begin{minipage}{8cm}
                \includegraphics[width=8.7cm]{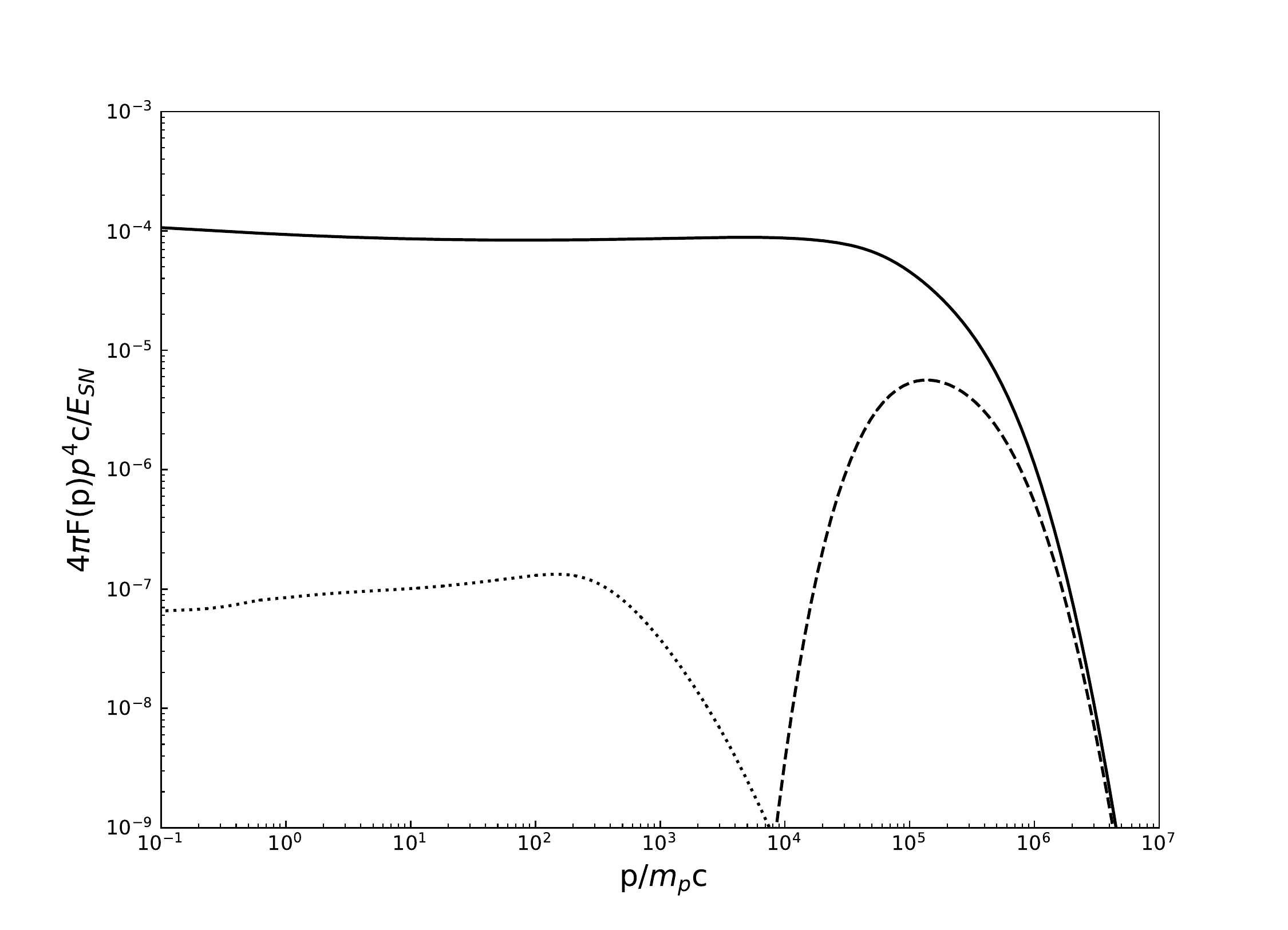}
        \end{minipage}
        \caption{Radial profiles of relevant physical quantities and spectra of particles produced in SNR. Top panel: Radial dependence of the gas pressure (solid line), the CR pressure (dotted line), and the magnetic field pressure (dashed line) at the epoch of $t=2000$ yr. Bottom panel: Spectra of particles produced in the whole SNR (space integrating) at the epoch of $t=2000$ yr. The spectra of protons include the intrinsic proton spectrum (solid line) and the electron spectrum (dotted line) inside the SNR, and escaped spectra (dashed line) integrated from $R_{\rm esc}=1.2R_f$ to $R_{\rm esc,out}= 10$ pc.}
        \label{fig:2}
\end{figure}

In this case, the changes in shock radii $R_f$ and $R_b$, the forward- and reverse-shock velocities $V_f$ and $V_b$ , and the ratio of CR energy to SN explosion energy ($E_{\rm cr}/E_{\rm sn}$) with time are shown in Fig. \ref {fig:1}. The calculations were performed until $t=2000\ \rm yr$, when the value of the forward-shock velocity drops down to $V_f=1384 \ \rm km\ s^{-1}$ and the forward-shock radius is $R_f=5.70\ \rm pc$. At the Sedov stage, the reverse-shock velocity quickly decreases to 78 $\rm km\ s^{-1}$ after $\sim 1000 $ years. Moreover, at 2000 yr, $E_{\rm cr}/E_{\rm sn}\approx 12\%$, approaching $E_{\rm cr}\sim 10^{49}$ erg. Considering the gas density $n_0=0.2\rm \ cm^{-3}$, the total CR energy is $\sim 3.0\times 10^{49} (n_0/0.2 \rm cm^{-3})^{-1}$, which can reproduce the $\sim \rm GeV$ gamma-ray observation \citep{Xin2019} for this source.

The profiles of the gas, CR pressure, and magnetic pressure at the radial distance at 2000 years are shown in the top panel of Fig. \ref{fig:2}, where the value of CR pressure is  much lower  than that of the gas pressure at the forward-shock position. The spatially integrated spectrum of accelerated protons and electrons is shown in the bottom panel of Fig. \ref{fig:2}. On the one hand, because of the severe synchrotron losses in strong magnetic field, the spectrum of the accelerated electrons appears to have an obvious break. They are cut off at a few TeV. On the other hand, the proton spectrum can be approximated as a power law with a spectral index of $\sim 2$ below a break energy $E_{\rm cut}$, and it decreases exponentially above $E_{\rm cut}$. The protons inside the SNR can be accelerated up to $\sim$ PeV, and the escaped protons mainly stem from the high-energy region with energy between $\sim E_{\rm cut}$ to $E_{\rm max}$ (see the dashed line in the bottom panel of Fig. \ref{fig:2}).

\begin{figure}
        \centering
%       \begin{minipage}{8cm}
%               \includegraphics[width=8.7cm]{G106sed_int.pdf}
%       \end{minipage}
        \begin{minipage}{8cm}
                \includegraphics[width=8.7cm]{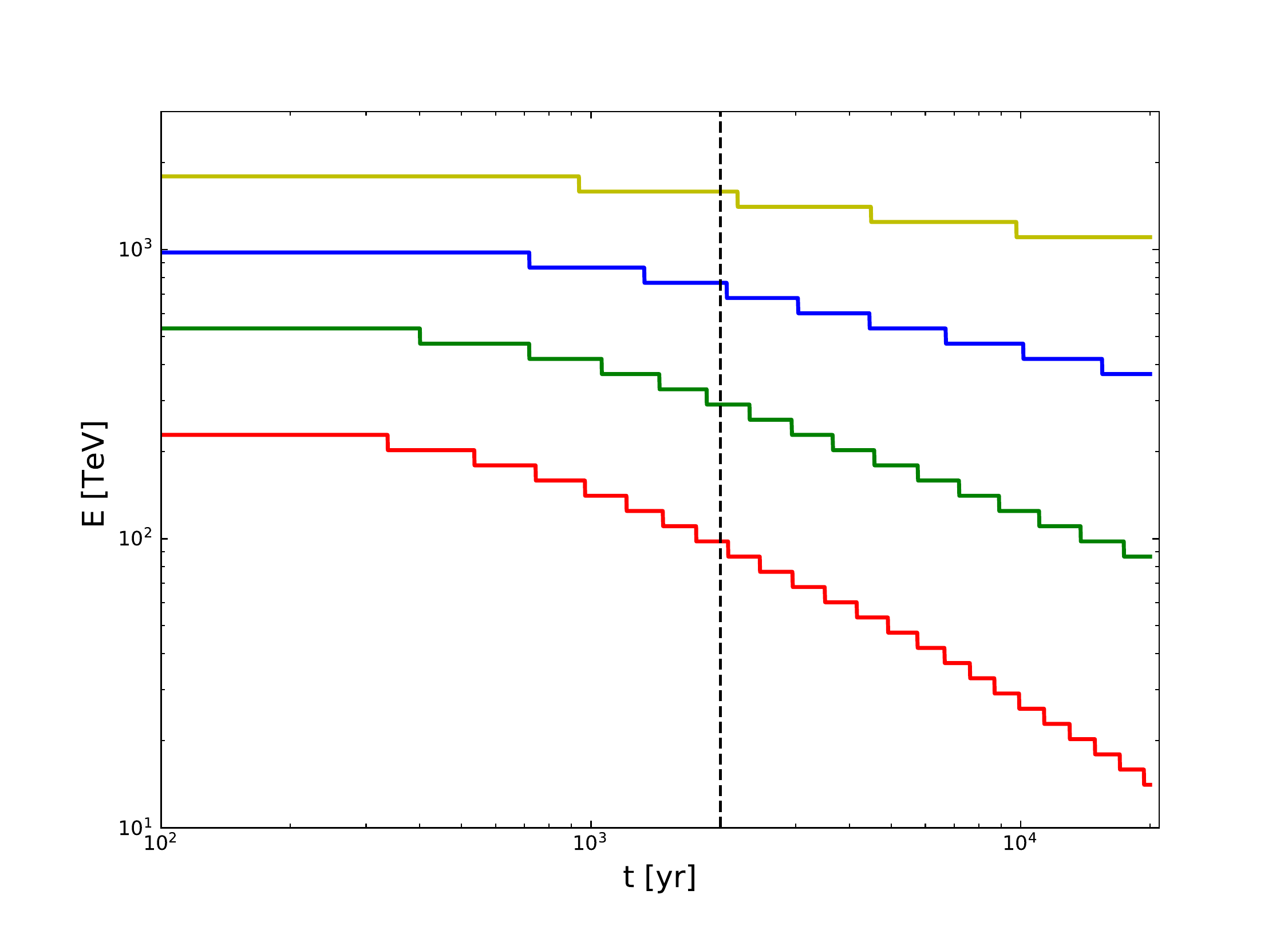}
        \end{minipage}
        \caption{%Top panel: The instantaneous spectra of accelerated protons (solid line) at the position of FS and escape protons (dashed line)  at age of 200yr (green) , 2000yr (red) and 20000 yr (blue) respectively. Bottom panel:
 Change of the energy $E$ of accelerated particles with time for $f=1$ (red line), $f=0.5$ (green line), $f=0.1$ (blue line), and $f=0.01$ (yellow line). The definition of $f$ is described in text.}
        \label{fig:3}
\end{figure}

For the maximum energy $E_{\rm max}$ of the accelerated protons, a simple estimate is as follows. Because of the free-escape boundary at the $R_{\rm esc}=\xi R_f$, maximum energy particles with distances greater than $R_{\rm esc}$ cannot diffuse back to the shock, and then the instantaneous maximum energy can be estimated by using $D(p_{\rm max})/V_f=(\xi-1) R_f$ \citep[e.g.,][]{Caprioli2010,Morlino2012}, where $D(p)$ is the diffusion coefficient. Here $\xi=1.2$ is used, which corresponds to a diffusing distance of $0.2 R_f$ upstream of the shock. The maximum energy for the Bohm-like diffusion is then approximated as
\begin{equation}
E_{\rm max}=10.4 \left(\frac{R_f}{{\rm pc}}\right)\left(\frac{V_f}{1000~{\rm km~s^{-1}}}\right)\left(\frac{B}{\mu G}\right)\;\;{\rm TeV}\;.
\end{equation}
 At the age of 2000 yr in our calculations, $R_f\sim5.7$ pc and $V_f\sim1384$ km s$^{-1}$, and for the instantaneous maximum magnetic field $B\sim50~\mu\rm G$ in the FS upstream, the estimated $E_{\rm max} \sim 2$ PeV. The magnetic field strength would fall back to the background magnetic field ($5\mu$G) as it is away from the immediate shock front (see the top panel of Fig. \ref{fig:2}). Therefore, the value of $E_{\rm max}$ estimated in Eq. 4 is an upper limit. On the other hand, the value of $E_{\rm max}$ can be roughly estimated as follows. Because the escape particles consist of those with maximum energies in different shock positions, their spectrum has a Gauss-like distribution, as shown in Fig . \ref{fig:2}. If $N_{\rm esc}(p_{\rm peak})$ represents the peak number density of the escape particles with $p_{\rm peak}$, then we can define a parameter $f=N_{\rm esc}(p\geqslant p_{\rm peak})/N_{\rm esc}(p_{\rm peak})$. For a given value of $f$, the change in corresponding $p$ with time can be calculated. The results for $f=1,~0.5,~0.1,$ and $0.01$ are shown in Fig.\ref{fig:3}, and the maximum energy $E_{\rm max}=p_{\rm max}c$ can reach $\sim$ PeV when $f\approx 0.01$. Therefore, the maximum energy of accelerated particles is $\sim 1 - \sim 2$ PeV. The values of the resulting step shape are only approximate because the momentum grid numbers are far lower than the time steps in our calculations.

%We also simulated the instantaneous particle spectra (including accelerating particles and escape particle) at the age of 200 yr, 2000yr and 20000 yr in the top panel of Fig.\ref{fig:3}, which indicate the escape particles spectra approximated to satisfied a. Supposing that  the maximum energy of $E_{max}=p_{max} c$ is the energy corresponding to, the evolution of $E_{max}$ is shown in bottom of, we can obtain the $E_{max}\sim 90$ TeV at the age of 2000 yr corresponding to  the cutoff energy ( $\sim 10$ TeV ) of TeV gamma ray emission inside the SNR.

%{\bf Regard  the maximum energy of observed gamma ray up to few hundred TeV \citep{Amene2021,Cao2021}, which would be produced by the interaction of the high energy particle of the early stage with MC (undergo the transportation in the Galaxy). Above descriptions the escape particles spectra satisfied with a distribution,  we could assume a parameter of $f=N(p\geqslant p_{max})/N_p(p_{max})$, where $N(p\geqslant p_{max})$ is the number density distribution of particles at the momentum of $p\geqslant p_{max}$.   If $f=1$, the $p_{max}$ could be obtained (see the above discussion). For other value of $f$ also shown in the bottom of Fig.\ref{fig:3}, the results show that the possible energy of $E$ of escape particles can be reach $\sim$ PeV when $f\geqslant 0.01$.}

 In our calculations, a low injection efficiency ($\eta_f=\eta_b=10^{-4}$) is adopted to reproduce the radio and GeV observation. If the injection efficiency for protons is high (e.g., $\eta_f=\eta_b=10^{-2}$), then the $E_{\rm cr}/E_{\rm sn}$ can be approximately $70\%$ after 1000 yr in ZP2012, and the total energy of the accelerated protons would be close to the energy of a supernova explosion. This high ratio ($E_{\rm cr}/E_{\rm sn} \sim 70\%$)  is higher than our result here ($E_{\rm cr}/E_{\rm sn}=10\%$). As a consequence, the produced GeV flux would exceed the observed flux for this source.

 To characterize the regions in which the escaped protons interact with the MC in the shell with a solid angle of $\Omega$ and a radius from $R_{\rm esc}$ to $R_{\rm esc,out}$ (the accelerated CR particles begin to run away from SNRs at the boundary ($R_{\rm esc}=\xi R_f$ )), we set the value of $R_{\rm esc,out}$ as $10$ pc to match the high-energy gamma ray observation. We introduced a parameter of $n_{\rm cloud}\Omega/4\pi=50\rm cm^{-3}$ to describe the number density of the MC, where the $n_{\rm cloud}$ is the number density of the MC. It is worth mentioning that $n_{\rm cloud}$ cannot be determined separately because of the relation to the size of radiation region (e.g., $\Omega=\pi/3$, $n_{\rm cloud}\sim 1500 \rm cm^{-3}$; $\Omega=4\pi$, $n_{\rm cloud}\sim 50 \rm cm^{-3}$). In addition, its typical density is about $100 \rm \ cm^{-3}$, wherein the cloud could host clumps with a density of $10^3-10^4 \rm \ cm^{-3}$ \citep{Owen2021}. The total mass of the MC in the interacting region is $\sim \frac{5\times 10^4}{\Omega}$ $M_\odot$ by integrating the density from $1.2 R_{\rm esc }$ to 10 pc.

% therefore the hadronic VHE emission luminosity is proportional to $\Omega n_{\rm cloud}$. The value of other parameters have been described in above section.

\begin{figure}
        \centering

                \includegraphics[width=8.7cm]{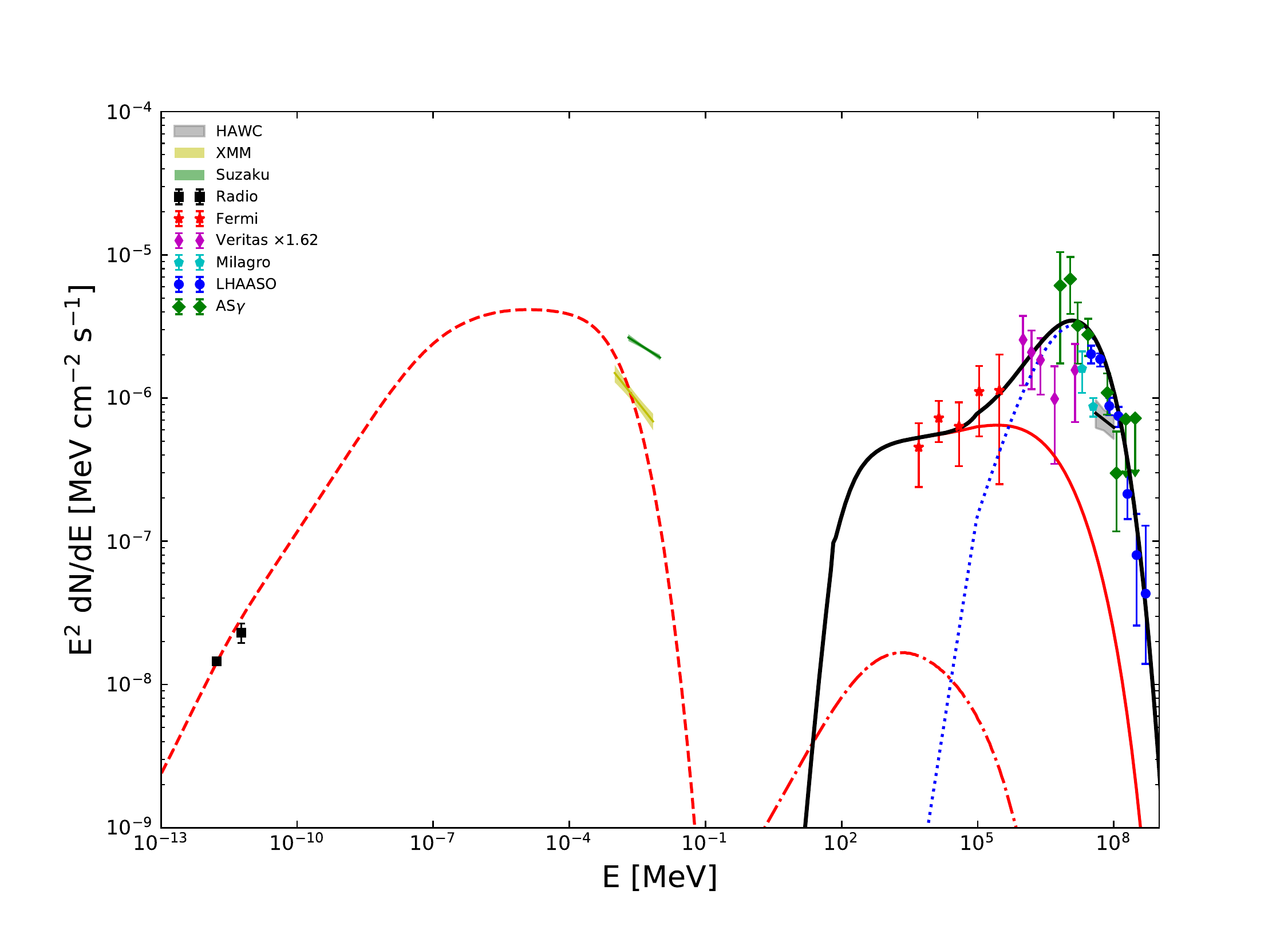}

        \caption{Multiwave band spectra of G106.3+2.7. The red line shows synchrotron (dashed line), inverse Compton (dot-dashed line), and pp interaction (solid line) inside the SNR.  The dotted blue line represents the high-energy emission from the escaped protons interacting with the MC, and the black line shows the overall VHE emission (pp interaction). Radio data of the tail region are from \cite{Pineault2000}. The statistical uncertainties of the tail region of XMM-Newton are from \cite{Ge2020}, and of the whole region of Suzaku, they are from \cite{Fujita2021}. The Fermi-LAT data are from \cite{Xin2019}, the VERITAS data are from \cite{Acciari2009}, the Milagro data are from \cite{Abdo2007,Abdo2009}, the statistical uncertainties of HAWC are from  \cite{Albert2020}, the AS$\gamma$ data are from \cite{Amene2021}, and the LHAASO data are from \cite{Cao2021} . }
        \label{fig:4}
\end{figure}

\begin{figure}
        \centering
        \begin{minipage}{8cm}
                \includegraphics[width=8.7cm]{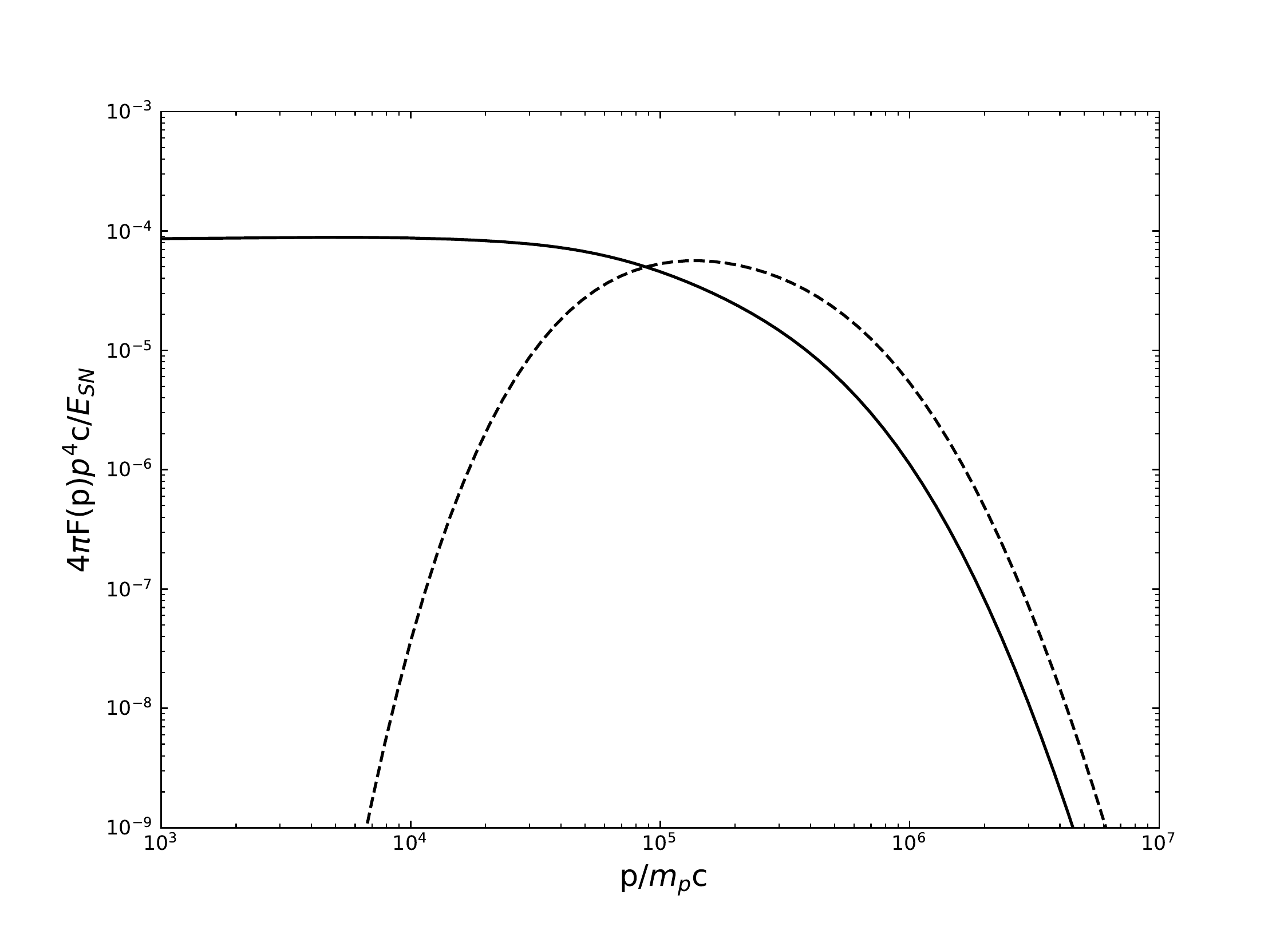}
        \end{minipage}
        \begin{minipage}{8cm}
                \includegraphics[width=8.7cm]{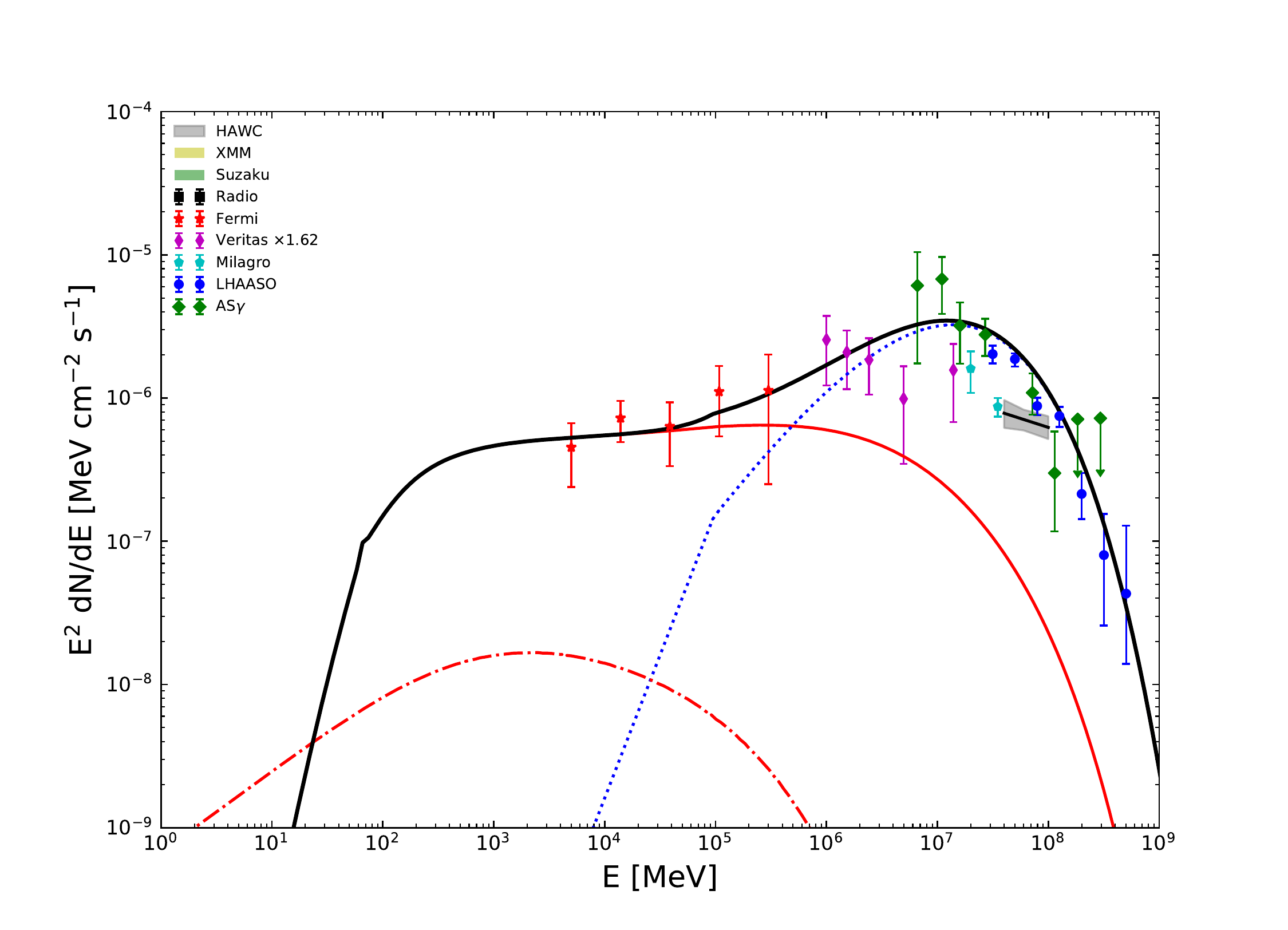}
        \end{minipage}
        \caption{Spectra of particles produced in SNR G106.3+2.7 as well as the multiwave band spectra. Top panel:  Same as the bottom panel of Fig.\ref{fig:2}, but the value of the escape particle spectra is magnified by an order of magnitude. Bottom panel: Same as Fig.\ref{fig:4}.}
        \label{fig:5}
\end{figure}

The particle (both the leptonic and hadronic) acceleration can be calculated by coupling hydrodynamic process and the evolution of the magnetic field (including the feedback between the magnetic field and and the accelerated protons)  inside the SNR,  as well as the extension to the escaped particle diffusion outside the SNR \citep{Tang2015}. For a given SNR, the spectra of accelerated particles and escaped particles can therefore be calculated naturally at any time $t$. Generally, the radio and X-ray photons originate from the leptonic synchrotron emission, and the high-energy gamma-rays are produced by the proton-proton interaction process (hadronic scenario) or the inverse-Compton (IC) process (leptonic scenario). Here the seed photons consist of 2.7 K microwave background light with an energy density of $0.25\rm eV cm^{-3}$, and a 25 K far-infrared radiation field with an energy density of $0.2 \rm eV cm^{-3}$ . Specifically, we used the emissivity formula developed by \cite{Blumenthal1970} for the leptonic component, while the proton-proton (pp) interaction process can be calculated by analytic approximation of \cite{Kelner2006}, where the hadronic scenario contains the intrinsic emission inside an SNR and the emission from its escaped protons colliding with a nearby MC.

The calculated spectral energy distribution (SED) of G106.3+2.7 with an estimated age of 2000 yr is shown in Fig. \ref{fig:4}. From radio to X-ray band, the photons are produced through the synchrotron emission of energetic electrons in the magnetic field with a strength of $\sim 50 \mu \rm G$ at the FS position (also see the evolution of magnetic field of Fig \ref{fig:1}). Thus, the resulting IC flux is very low due to severe radiative cooling in the strong magnetic field. \citet{Xin2019} also argued that the radio flux gives a magnetic field strength of $\sim 50\mu \rm G$ and a corresponding $E_{\rm cut}\sim$ TeV of the accelerated electrons in a hadronic scenario. The gamma-ray emission mainly originates from the hadronic component, and the Fermi data can be fit by the accelerated protons inside the SNR together with the escaped protons outside the SNR interacting with the nearby MC. A clearer picture is presented in Fig.\ref{fig:5}. Our results indicate that the spectrum of high-energy gamma-rays is dominated by the hadronic component, and the overall gamma-ray flux can naturally be explained by our model.

\section{Summary and discussion}

The multiband photon emission from SNR G106.3+2.7 (a potential Galactic PeVatron) was studied here. The results show that photons with energy of $E_\gamma \gtrsim $ 1 GeV favor a hadronic origin, where the photons in the energy range of  $\sim 1$ GeV to $\sim 1$ TeV are produced inside the SNR through proton-proton interaction, and photons with $E_\gamma \gtrsim 1$ TeV originate from the interaction of escaped protons with a dense molecular cloud.

 In our model, the magnetic field plays a dynamical role. Magnetic pressure and magnetic energy flux are taken into account in the shock. The amplified effects and the spatial distribution of the magnetic field were assumed using a simple parameterized description in the following.  The parameter $M_A$ and the plasma density determine the value of the amplified magnetic field strength for $r > R_c$ (see Eq. 1). In the vicinity of the CD, the plasma flow is strongly influenced by the Rayleigh-Taylor instability and results in the generation of magnetohydrodynamic (MHD) turbulence in this region. Eq. 2 is assumed to describe the amplification of the downstream magnetic field of the reverse shock. However, one important mechanism for generating an amplified magnetic field is the plasma instabilities driven by CR streaming \citep{Bell2013,Caprioli2009}, which in turn leads to a more efficient particle acceleration. In particular, the shock permeating into the dense medium is able to excite nonresonant streaming instabilities, and efficiently accelerates particles to the PeV range in the early stages \citep{Cristofari2020a}. It is noteworthy that the nonresonant growth of instabilities is expected to lead to the PeV range in only very few cases \citep{Bell2013,Schure2013,Cristofari2020b}. For instance, based on the analysis of the nonresonant instability, \citet{Schure2013} reported that, in the context of SNR shocks, only young SNRs in a dense environment are plausible candidates to accelerate CRs to PeV energies under some conditions (see also \cite{Cristofari2020b} for a more recent discussion). Therefore, a further investigation of MHD effects at astrophysical shocks is necessary.

\begin{figure}
        \centering

                \includegraphics[width=8.7cm]{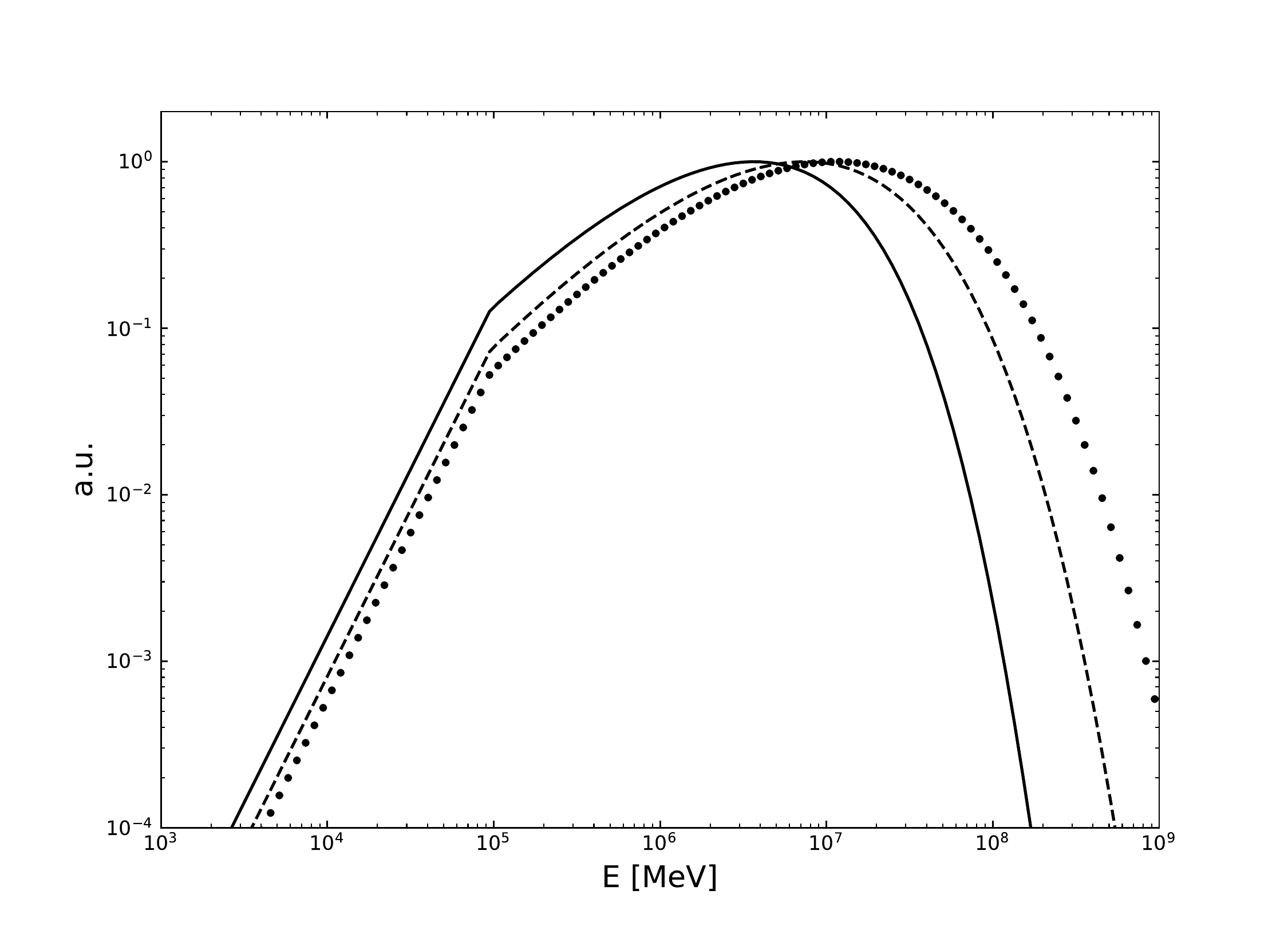}

        \caption{Interaction of gamma-rays from the escaped protons with the same density of the MC for $\chi=0.1$ (solid line), $0.01$ (dashed line), and $0.001$ (dotted line),  where the gamma-ray flux maximum value is normalized to 1.}
        \label{fig:6}
\end{figure}

 In the escaping region, parameter $\chi$ of the diffusion coefficient in Eq. 3 is taken as $0.001$ to fit the TeV gamma-ray spectra observed by LHAASO,  which is much lower than its counterpart in the Galaxy. Near the SNR, the diffusion coefficient can be suppressed by a factor of 100 or more compared to the Galaxy \citep{Fujita2009}. In addition, we calculated the gamma-ray emission spectra for different $\chi$ values (see Fig. \ref{fig:6}). The results indicate that the diffusion coefficient greatly affects the high-energy tail of gamma-rays.

 We set the age of the SNR to 2000 yr to reproduce the high-energy data. Fig. \ref{fig:3} shows that the maximum energy of acceleration particles decreases rapidly at a late stage, and an older age (e.g., $10^4$ yr) makes it harder to fit the high-energy data. On the other hand, the energetic escaped particles accelerated at an early stage should not reside in the interacting region with the MC.

 The typical explosion energy for a thermonuclear supernova is $E_{\rm sn}=10^{51}$ erg. Regarding the age of SNR G106.3+2.7 (2000 yr) and the discussion in Sect. \ref{per}, the ambient medium number density is about $2\ \rm cm^{-3}$. The total CR energy is $\sim 3.0\times 10^{48} (n_0/2.0 \rm cm^{-3})^{-1}$, which leads to $E_{\rm cr}/E_{\rm sn}\sim0.1\%$ for a medium number density of $2\ \rm cm^{-3}$ with an explosion energy $E_{\rm sn}=10^{51}$ erg. Therefore, the insufficient acceleration of CR protons in the shock is invoked. Everything we know about DSA suggests that young, strong SNR shocks are intrinsically much more efficient than our case here, for instance, $E_{\rm cr}/E_{\rm sn} \sim 10\%$. It is very difficult to produce TeV photons via DSA with this low acceleration efficiency. Hence, the explosion energy adopted here is $E_{\rm sn}=10^{50}$ erg, while the estimated  ambient medium number density is $n_0\sim0.2$ cm$^{-3}$ \citep{Kothes2001}.

 Here, the evolution and particle acceleration only consider type Ia SNR. However, ignoring other types is probably too simple for this specific SNR, as the actual types of the parent supernova explosion and the progenitor star are still undetermined. For type Ia SNRs, the matter distribution in the circumstellar medium (CSM) and interstellar medium (ISM) is less affected by the activities of the progenitor stars, which may result in a statistically more spherical shape. For core-collapse SNRs, however, the stellar winds blown by progenitor stars would have modified the CSM and ISM matter distribution prior to the explosion. As a consequence, the hadronic processes are affected as the matter distribution varies. A more detailed numerical simulation of particle acceleration in four types of SNRs has been conducted by \cite{Ptuskin2010}. Their results indicated that the maximum energy of accelerated particles in all stages can be different for various types.

 \citet{Bao2021} have studied the hard $\gamma$-ray SED of SNR G106.3+2.7 and pointed out that the radio to X-ray band and the Fermi gamma-rays can originate from the synchrotron emission and inverse Compton of the energetic electrons, respectively. Above 10 TeV, gamma-ray emission arises from the contribution of escaped particles that interact with the MC. However, the differences between their model and our model are as follows. The acceleration and propagation of the particles are self-consistently considered through the ZP2012 model in the acceleration region with its extension \citep{Tang2015} to escaping regions, and the hadronic contribution from two regions is predicted in our model. Moreover, the resulting IC flux with $E_\gamma\gtrsim 100$ MeV is much lower than that produced in proton-proton interaction inside the SNR because of the energetic electron cooling in a strong magnetic field.

Finally, although the gamma-ray emission can be produced via high-energy particles accelerated by the shock of the SNR  G106.3+2.7 \citep{Albert2020,Liu2020}, with hadronic processes dominating the photon spectrum extending to at least 400 TeV \citep[]{Xin2019,Amene2021}, it should be pointed out that a leptonic origin of VHE $\gamma$-rays is also possible, such as in pulsar wind nebulae \citep{Albert2020}. Therefore, deeper observations are required.

%%%%%%%%%%%%%%%%%%%%%%%%%%%%%%%%%%%%%%%%%%%%%%%%%%%%%%%%%%%%%%%%%%%%%%%%%%%%%%%%%%%%%%%%%%%%%%%%%%%%%%%%%%%%%%%%%%%%%%%%%%%%%%

%%%%%%%%%%%%%%%%%%%%%%%%%%%%%%%%%%%%%%%%%%%%%%%%%%%%%%%%%%%%%%%%%%%%%%%%%%%%%%%%%%%%%%%%%%%%%%%%%%%%%%%%%%%%%%%%%%%%%%%%%%%%%%

\begin{acknowledgements}
This work is partially supported by National Key R \& D Program of China under grant No. 2018YFA0404204, and the National Natural Science Foundation of China U1738211. C.Y. Yang is partially supported by the National Natural Science Foundation of China U2031111, U1931204, 11673060.
    \end{acknowledgements}

%%%%%%%%%%%%%%%%%%%%%%%%%%%%%%%%%%%%%%%%%%%%%%%%%%%%%%%%%%%%%%%%%%%%%%%%%%%%%%%%%%%%%%%%%%%%%%%%%%%%%%%%%%%%%%%%%%%%%%%%%%%%%%

%%%%%%%%%%%%%%%%%%%%%%%%%%%%%%%%%%%%%%%%%%%%%%%%%%%%%%%%%%%%%%%%%%%%%%%%%%%%%%%%%%%%%%%%%%%%%%%%%%

\end{document}